# Preparation and Determination of Spin-Polarized States in Multi-Zeeman-sublevel Atoms

Bo Wang,[1] Yanxu Han,[1] Jintao Xiao,[1] Xudong Yang,[1] Chunhong Zhang,[1] Hai Wang,[1,*] Min Xiao[1,2], and Kunchi Peng[1]

[1] *The State Key Laboratory of Quantum Optics and Quantum Optics Devices, Institute of Opto-Electronics, Shanxi University, Taiyuan, 030006, People's Republic of China*

[2] *Department of Physics, University of Arkansas, Fayetteville, Arkansas 72701, USA*

We demonstrate a simple technique to prepare and determine the desired internal quantum states in multi-Zeeman-sublevel atoms. By choosing appropriate coupling and pumping laser beams, atoms can be easily prepared in a desired Zeeman sublevel with high purity or in any chosen ground-state population distributions. The population distributions or state purities of such prepared atomic states can be determined by using a weak, circularly-polarized probe beam due to differences in transition strengths among different Zeeman sublevels. Preparing well-defined internal quantum states in multi-Zeeman-sublevel atoms (or spin-polarized quantum-state engineering) will be very important in demonstrating many interesting effects in quantum information processing with multi-level atomic systems.

Preparing atoms into one specified internal quantum state and determining the population distribution in multi-Zeeman-sublevel atomic systems are very important in studying atom-field interactions, especially interesting schemes for quantum information processing such as light storage [1], quantum phase gate [2-5], and entanglement between atomic assemble and photons [6] or between a single trapped ion and a single photon [7]. Demonstrations of these novel effects require more than two atomic energy levels and well-defined initial internal quantum state for the atoms, which can not be accomplished by simple optical pumping as in the case for a two-level atomic system.  Although in most cases interesting effects can be experimentally demonstrated by simply considering degenerate Zeeman levels, so no specific ground-state population preparations are needed (as in the cases of electromagnetically induced transparency (EIT) [8-10] and photon storage [11]), there are many effects that demand better quantum-state preparation and determination in the multi-Zeeman-sublevel atomic systems.  For example, in order to demonstrate quantum phase gate in multi-level atomic systems, such as the five-level M-type [3] and five-level combined M and tripod-type [5] systems, initial ground-state populations have to be prepared in specific Zeeman sublevels. Other examples include synthesis of arbitrary quantum states [12] and many other multi-level atomic systems for quantum information processing.  Although specific atomic ground states were prepared in some of the previous experiments and the population distributions were estimated [6], no simple optical techniques have been developed, to the best of our knowledge, to determine the ground-state populations of the prepared internal quantum states of the atoms. Such measurements are very important in determining the coherent time of the photon storage [13], achieving large entanglement between photons and the atomic assemble [6], and realizing quantum phase gates due to cross-phase modulation [3-5] in the multi-level atomic assembles. Previously demonstrated techniques to measure ground-state population distributions include using a set of Stern-Gerlach measurements [14] and making use of light-induced momentum transfer [15,16].

In this letter, we present a simple procedure to prepare any desired internal quantum states at various Zeeman sublevels, and more importantly, we demonstrate a new technique to determine the ground-state population distributions simply by applying a weak circularly-polarized probe laser beam. The preparation of the desired quantum state is achieved by combining optical pumping from two polarized laser beams. The basic mechanism for detecting and determining the prepared internal quantum state is to make use of the differences in transition strengths among different Zeeman sublevels (different Clebsch-Gordan coefficients), as shown in Fig. 1(b) for the D1 line of $^{87}$Rb atoms, and the changes in the corresponding multi-dark-state resonances (MDSR) or multiple EIT structure [as shown in Fig. 1(c)] [17]. By combining the measured probe spectrum with the theoretical calculation, one can easily determine the ground-state population distribution and the purity of the prepared quantum state with high precision.

The experiment was performed with D1 line of $^{87}$Rb atoms, as shown in Fig. 1(b), in a vapor cell magneto-optical trap (MOT) [17]. The diameter of the trapping beam is about 1 cm. The trapped $^{87}$Rb atom cloud is about 2 mm in diameter and contains about $10^9$ atoms at the temperature of about 200 μK. The probe and coupling beams are from two extended cavity diode lasers with linewidth of about 2 MHz. The probe laser frequency is scanned at a speed of 1.3 kHz/s across the $5S_{1/2}$, F=1 → $5P_{1/2}$, F'=2 transition and the frequency of the coupling laser is locked to the $5S_{1/2}$, F=2 → $5P_{1/2}$, F'=2 transition. The weak probe beam is left-circularly-polarized and the strong coupling beam is linearly polarized in z direction [as shown in Fig. 1(a)]. The coupling beam propagates along x axis through the cold atoms while the probe beam propagates along z axis through the cold atoms. The diameters of the coupling beam and the probe beam are 2.6 mm and 0.5 mm, respectively. A uniform magnetic field of about 150 mG in the direction along z axis is applied at the location of the cold atoms. Two beams overlap at the center of the cold atom cloud. During the experiment, the on-off sequence of the trapping, repumping, coupling, and probe beams is controlled by four acousto-optical modulators (AOM). The experiment is cycled at 10 Hz. In each cycle, first 99 ms is used for the cooling and trapping of the $^{87}$Rb atoms. We switch on the coupling beam at the same time as we turn off the MOT (trapping and repumping beams, as well as the magnetic field). 0.1 ms later, the probe beam is turned on and the transmitted probe beam from the cold atoms is detected by an avalanche photodiode.

The relevant energy levels are shown in Fig. 1(b) with a linearly-polarized coupling beam $E_c$ (frequency $\omega_c$) drives transition 5S$_{1/2}$, F=2 to 5P$_{1/2}$, F'=2. Since the transition strength between states $|b_{-2}\rangle$ and $|c_{-2}\rangle$ is twice the value between states $|b_{-1}\rangle$ and $|c_{-1}\rangle$, the coupling beam Rabi frequency $\Omega_{c2}$ (connecting states $|b_{-2}\rangle$ and $|c_{-2}\rangle$, defined as $\frac{\mu_{b_{-2}c_{-2}} E_c}{\hbar}$, where $\mu_{b_{-2}c_{-2}}$ is the dipole moment between the two states) is twice the value of $\Omega_{c1}$ (connecting the states $|b_{-1}\rangle$ and $|c_{-1}\rangle$). A left-circularly-polarized probe beam $E_p$ (frequency $\omega_p$) scans through the transition 5S$_{1/2}$, F=1 to 5P$_{1/2}$, F'=2. The probe field $\Omega_{p2}$ (connecting states $|a_{-1}\rangle$ and $|c_{-2}\rangle$, defined as $\frac{\mu_{a_{-1}c_{-2}} E_p}{\hbar}$, which is about 1 MHz in the experiment) and the coupling field $\Omega_{c2}$ (about 78 MHz in the experiment) form an EIT system, which has an EIT window much broader than the EIT window formed by the probe field $\Omega_{p1}$ (connecting states $|a_0\rangle$ and $|c_{-1}\rangle$) and the coupling field $\Omega_{c1}$ ($\Omega_{c1}=\Omega_{c2}/2$). Since the transition strength between states $|b_0\rangle$ and $|c_0\rangle$ is zero [18], the probe field $\Omega_{p0}$ (connecting states $|a_{+1}\rangle$ and $|c_0\rangle$) has a simple absorption peak. Note that $\Omega_{p0}$, $\Omega_{p1}$, and $\Omega_{p2}$ are Rabi frequencies for the same probe laser beam $E_p$ corresponding to different transitions and similarly, $\Omega_{c1}$ and $\Omega_{c2}$ are the Rabi frequencies for the same coupling field $E_c$ corresponding to different transitions. When the probe beam scans through the transition from $5S_{1/2}$, F=1 to $5P_{1/2}$, F'=2, one observes the multiple EIT peaks (or MDSR), as shown in Fig. 1(c), due to the multiple EIT windows plus the central absorption peak [17].

To quantitatively determine the ground-state populations of the prepared atomic spin states, we theoretically modeled this multi-Zeeman-sublevel atomic system with 13 atomic energy levels and two laser beams. The susceptibilities of the probe field coupling the states $|a_{-1}\rangle \to |c_{-2}\rangle$, $|a_0\rangle \to |c_{-1}\rangle$, and $|a_{+1}\rangle \to |c_0\rangle$ are:

$$\chi_{a_{-1}c_{-2}} = -\frac{P_{a_{-1}} N_{F1}}{\hbar \varepsilon_0} \cdot \frac{|\mu_{a_{-1}c_{-2}}|^2}{\Omega_{p2}} \rho_{a_{-1}c_{-2}},$$

$$\chi_{a_0 c_{-1}} = -\frac{P_{a_0} N_{F1}}{\hbar \varepsilon_0} \cdot \frac{|\mu_{a_0 c_{-1}}|^2}{\Omega_{p1}} \rho_{a_0 c_{-1}}, \qquad (1)$$

$$\chi_{a_{+1}c_0} = -\frac{P_{a_{+1}} N_{F1}}{\hbar \varepsilon_0} \cdot \frac{|\mu_{a_{+1}c_0}|^2}{\Omega_{p0}} \rho_{a_{+1}c_0},$$

where $P_{a_{-1}}$, $P_{a_0}$, and $P_{a_{+1}}$ are the populations of the states $|a_{-1}\rangle$, $|a_0\rangle$, and $|a_{+1}\rangle$, respectively. $N_{F1}$ is the total density of atoms staying in the energy levels of $5S_{1/2}$, F=1. The density-matrix elements $\rho_{a_{-1}c_{-2}}$, $\rho_{a_0 c_{-1}}$, and $\rho_{a_{+1}c_0}$ can be calculated by solving the density-matrix equations involving all Zeeman sublevels in the system [19]. The total probe field susceptibility is given by

$$\chi = \chi_{a_{-1},c_{-2}} + \chi_{a_0,c_{-1}} + \chi_{a_{+1},c_0}. \qquad (2)$$

By fitting numerically calculated curves from Eq. (2) to the measured MDSR signal, as shown in Fig. 1(c), one can determine the population distributions $P_{a_{-1}}$, $P_{a_0}$, and $P_{a_{+1}}$. We first fit the theoretically calculated results (dotted line) to the experimentally measured data (solid line) in Fig.1(c) and find excellent match between them. From such best fitting, the ground-state population distribution in Zeeman sublevels m=-1, 0, +1 of $5S_{1/2}$, F=1 level is determined to be $P_{a_{-1}} = 32\%$, $P_{a_0} = 36\%$, and $P_{a_{+1}} = 32\%$.

To prepare atoms into a well-specified spin-polarized state, a polarized pumping beam $E_b$ is added, which is on resonance with the transition from $5S_{1/2}$, F=1 to $5P_{1/2}$, F'=1 and propagates through the cold atoms with a small angle (about 2°) relative to the probe beam in the same plane [as shown in Figs. 2(a1) and (a2)]. The power of the pumping beam is set at 13.6 mW with a diameter of ~ 2 mm (which overlaps with the coupling and probe beams at the cold atomic cloud). The experimental procedure is the same as the one discussed above by switching on the coupling beam and this pumping laser beam at the same time as we turn off the MOT.

When the pumping beam is left-circularly polarized (dashed lines), as shown in Figs. 2(a1) and (b1), all the ground-state populations in $5S_{1/2}$, F=1 are prepared in the m=-1 ($|a_{-1}\rangle$) state. In this case, when the weak, left-circularly-polarized probe beam (dotted line) scans through the transition, only one simple EIT curve (related to the energy levels $|a_{-1}\rangle$, $|c_{-2}\rangle$, and $|b_{-2}\rangle$) is observed, as shown in Fig. 3(a) (solid curve). The two absorption peaks correspond to the dressed-state absorption due to the coupling field $\Omega_{c2}$. The clean double-peak structure is a good indication that atoms are mostly pumped into the $|a_{-1}\rangle$ state. Similarly, when the pumping beam is right-circularly polarized, as shown in Figs. 2(a2) and (b2), all the ground-state populations in $5S_{1/2}$, F=1 are prepared in the m=+1 ($|a_{+1}\rangle$) state. In this case, since no coupling between states $|b_0\rangle$ and $|c_0\rangle$, no EIT system can be formed and only a single absorption peak is observed, as shown in Fig. 3(b) (solid line). As the pumping beam is changed to be linearly polarized, as shown in Figs. 2(a3) and (b3) (dashed lines), all the ground-state populations in $5S_{1/2}$, F=1 are prepared in the m=0 ($|a_0\rangle$) state. Notice that the transition strength between states $5S_{1/2}$, F=1, m=0 and $5P_{1/2}$, F'=1, m'=0 is also zero [17]. In this case, the left-circularly-polarized probe field $\Omega_{p1}$ (dotted line) and the coupling field $\Omega_{c1}$ (connecting states $|b_{-1}\rangle$ and $|c_{-1}\rangle$) form a single EIT system, as shown in Figs. 2(b3) and 3(c). This EIT window is much narrower than the one for a left-circularly-polarized pumping beam [Fig. 3(a)], since the coupling Rabi frequency $\Omega_{c1}=\Omega_{c2}/2$. The clean single EIT curve indicates the concentration of ground-state population in state $|a_0\rangle$, since any residual populations in other states will give rise to peaks at exact resonance (due to absorption in the transition from $|a_{+1}\rangle$ to $|c_0\rangle$) and far sides due to larger EIT window formed by states $|a_{-1}\rangle$, $|c_{-2}\rangle$, and $|b_{-2}\rangle$.

Next, we add the pumping beam in the theoretical model and calculate the probe beam susceptibilities. By letting the pumping beam to be left-circularly-, right-circularly-, and linearly-polarized, respectively, in the theoretical calculations, we can calculate the probe beam susceptibilities and fit them to the experimentally measured results as given in Fig. 3.

The atomic decay rates are taken to be $\gamma_{ab}$=2 MHz and $\gamma_{ac}$=4 MHz (both of them include the laser linewidths). As one can see that excellent matches between the theoretical calculated results (dotted lines) and experimentally measured data (solid lines) are obtained, from which we can deduce the population distributions of the ground states or purities of the prepared quantum states of atoms. As the fitting indicates, the populations in the $5S_{1/2}$, F=1, m=-1, 0, +1 states are $P_{a_{-1}} = 96\%$, $P_{a_0} = 2\%$, $P_{a_{+1}} = 2\%$, respectively, with the total density of the $5S_{1/2}$, F=1 states $N_{F1}$=0.6×10$^{11}$ cm$^{-3}$ for the left-circularly-polarized pumping beam [as in Fig. 3(a)]. For the right-circularly-polarized pumping beam, the ground-state populations are $P_{a_{-1}} = 1\%$, $P_{a_0} = 1\%$, $P_{a_{+1}} = 98\%$, respectively, with $N_{F1}$=0.6×10$^{11}$ cm$^{-3}$ for the three ground states [as in Fig. 3(b)]. When the pumping beam is changed to be linearly polarized, the ground-state populations are $P_{a_{-1}} = 1\%$, $P_{a_0} = 98\%$, $P_{a_{+1}} = 1\%$, respectively, with $N_{F1}$=0.6×10$^{11}$ cm$^{-3}$ [as given in Fig. 3(c)]. In all these cases, the pumping power is kept to be a constant (13.6 mW). As one can see that the purities of the prepared spin-polarized states are quite high (>96%). It is worth to note that the total density of atoms in the $5S_{1/2}$, F=1 states is $N_{F1}$=1.22×10$^{11}$ cm$^{-3}$ in the absence of the pumping beam, while $N_{F1}$ becomes 0.6×10$^{11}$ cm$^{-3}$ in the presence of the polarized pumping beam with a power of about 13 mW. The total atomic population includes atomic populations in the Zeeman sublevels of $5S_{1/2}$, F=1 ($N_{F1}$) and $5S_{1/2}$, F=2 ($N_{F2}$) states. Since the state $5S_{1/2}$, F=2, m=0 ($|b_0\rangle$) does not interact with any other states in the schemes [as shown in Figs. 1(c) and 2(b1)-(b3)], certain atomic population will be trapped into this $|b_0\rangle$ state as $N_{F2}$. The ratio between $N_{F1}$ and $N_{F2}$ changes as the polarized pumping beam is turned on, which modifies the total ground-state population on the $5S_{1/2}$, F=1 state ($N_{F1}$).

Using this technique, we can not only prepare pure spin-polarized ground states in this multi-Zeeman-sublevel atomic system, but also create and determine any desired ground-state population distributions. By decreasing the intensity of the pumping beam in different pumping beam polarizations (as in the cases of Fig. 2), we can obtain different ground-state population distributions. Figure 4 shows different probe signals obtained at different powers for left-circularly-polarized pumping beam [Fig. 2(b1)] and the fitted results. When the pumping power is at 5 mW, a good fitting with the experimental data gives $P_{a_{-1}} = 92\%$, $P_{a_0} = 4\%$, and $P_{a_{+1}} = 4\%$, $N_{F1}=0.6\times10^{11}$ cm$^{-3}$ as shown in Fig. 4(a). As the pumping power is increased to be larger than 5 mW, atoms will mostly be pumped into the |a$_{-1}$> state, as the case shown in Fig. 3(a). When the pumping power is lowered to 1 mW, a ground-state population distribution of $P_{a_{-1}} = 64\%$, $P_{a_0} = 18\%$, $P_{a_{+1}} = 18\%$, as shown in Fig. 4(b), is obtained. As the pumping power is further reduced to 0.5 mW, the ground-state population distribution becomes $P_{a_{-1}} = 42\%$, $P_{a_0} = 33\%$, $P_{a_{+1}} = 25\%$, as shown in Fig. 4(c). When the pumping power is 50 μW, the ground-state population distribution is $P_{a_{-1}} = 33\%$, $P_{a_0} = 35\%$, $P_{a_{+1}} = 32\%$, as shown in Fig. 4(e), which is basically the same as the results when without the pumping beam [Fig. 1(c)]. Similarly, with right-circularly-polarized or linearly-polarized pumping beams at different powers, various other ground-state population distributions can be obtained. Actually, any desired ground-state population distributions can be realized by choosing appropriate pumping beam polarization and power, e.g. by entering the desired population distribution into the model, one can calculate the needed pumping beam power and polarization. This is a very powerful technique in preparing designed spin-polarized states in multi-Zeeman-sublevel atomic systems for using in quantum-state engineering.

In summary, we have developed an efficient and easy way to prepare desired spin-polarized ground-state population distribution in a multi-Zeeman-sublevel atomic system and demonstrated a simple optical scheme to determine the population distributions with high precision. This technique is based on the differences in the transition strengths among different Zeeman sublevels and the companion MDSR spectra.  We have shown that more than 96% state purity can be easily achieved in any one of the ground states, and any other desired state mixings can also be prepared and determined, which constitute a complete spin-polarized quantum state engineering. Such prepared initial atomic states can be used for demonstrating quantum phase gates in four- and five-level atomic systems, improving quantum state storage in atomic assembles, preparing atomic states for entanglement, and using in many other quantum information processing schemes.

We acknowledge funding supports by the National Natural Science Foundation of China (# 60325414, 60578059, 60238010).  *Corresponding author H. Wang's e-mail address is wanghai@sxu.edu.cn.

1. C. Liu, Z. Dutton, C. H. Behroozi, and L. V. Hau, Nature (London) **409**, 490 (2001).

2. Q. A. Turchette, C. J. Hood, W. Lange, H. Mabuchi, and H. J. Kimble, Phys. Rev. Lett. **75**, 4710 (1995).

3. C. Ottaviani, D. Vitali, M. Artoni, F. Cataliotti, and P. Tombesi, Phys Rev. Lett. **90**, 197902 (2003); C. Ottaviani, S. Rebić, D. Vitali, and P. Tombesi, Phys. Rev. A **73**, 010301(R) (2006).

4. X. B. Zou and W. Mathis, Phys. Rev. A **71,** 042334 (2005).

5. Z.-B. Wang, K.-P. Marzlin, and B. C. Sanders, Phys Rev. Lett. **97**, 063901 (2006).

6. D. N. Matsukevich, T. Chanelière, S. D. Jenkins, S.-Y. Lan, T. A. B. Kennedy, and A. Kuzmich, Phys. Rev. Lett. **96**, 030405 (2006).

7. B. B. Blinov, D. L. Moehring, L.-M. Duan, and C. Monroe, Nature **428**, 153 (2004).

8. S. E. Harris, Phys. Today **50**, 36 (1997).

9. K. J. Boller, A. Imamoglu, and S.E. Harris, Phys. Rev. Lett. **66**, 2593 (1991).

10. J. Gea-Banacloche, Y. Q. Li, S.-Z. Jin, and Min Xiao, Phys. Rev. A **51**, 576 (1995).

11. T. Chanelière, D. N. Matsukevich, S. D. Jenkins, S.-Y. Lan, T. A. B. Kennedy, and A. Kuzmich, Nature **438**, 833 (2005).

12. A. S. Parkins, P. Marte, P. Zoller, and H.J. Kimble, Phys. Rev. Lett. **71**, 3095 (1993).

13. D. N. Matsukevich, T. Chanelière, S. D. Jenkins, S.-Y. Lan, T. A. B. Kennedy, and A. Kuzmich, Phys. Rev. Lett. **96**, 033601 (2006).

14. G. S. Summy, B. T. H. Varcoe, W. R. MacGillivray, and M. C. Standage, J. Phys. B **30**, L541 (1997).

15. L. S. Goldner, C. Gerz, R. J. C. Spreeuw, S. L. Rolston, C. I. Westbrook, and W. D. Phillips, P. Marte, and P. Zoller, Phys. Rev. Lett. **72**, 997 (1994).

16. G. Klose, G. Smith, and P. S. Jessen, Phys. Rev. Lett. **86**, 4721 (2001).

17. Bo Wang, Y. Han, J. Xiao, X.Yang, C. D. Xie, H. Wang, and Min Xiao, Opt. Lett., doc. ID 74036 (posted 28 September 2006, in press).

18. For details of the transition probabilities of D1 line in $^{87}$Rb, see http://steck.us/alkalidata.

19. S. Li, Bo Wang, X. Yang, Y. Han, H. Wang, Min Xiao, and K. C. Peng, Phys. Rev. A **74**, 033821 (2006).

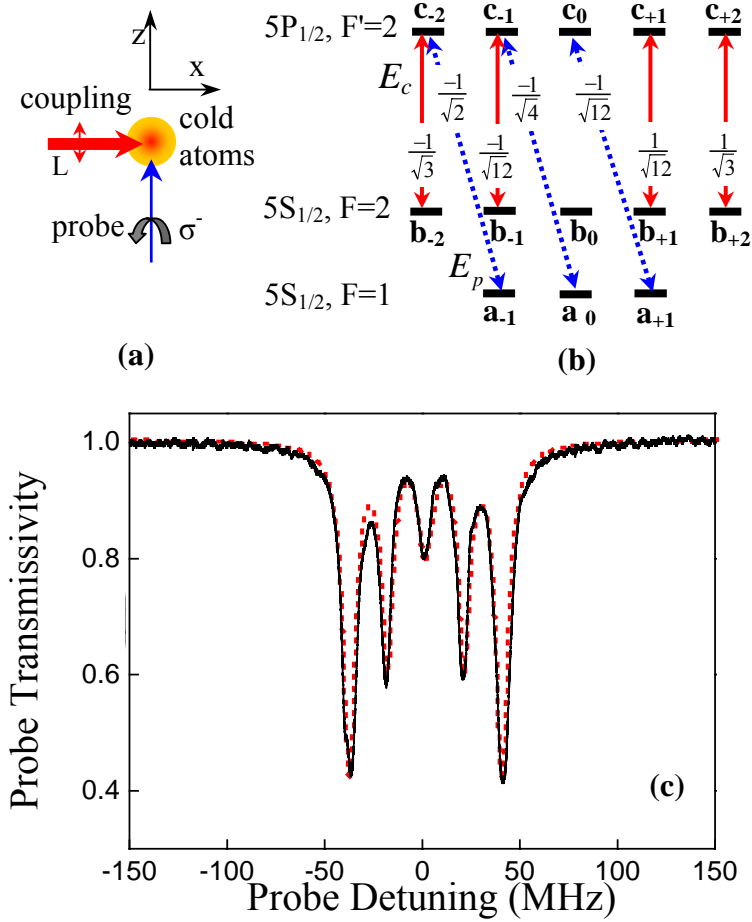

FIG. 1. (Color online) (a) Experimental setup. L: linearly polarized, σ⁻: left-circularly polarized. Z: the direction of uniform magnetic field (~150 mG). (b) Relevant atomic energy levels interacting with the probe (dotted lines) and coupling (solid lines) laser beams. (c) Transmission spectra of the probe beam with $\Omega_{c2}$=78 MHz and $\Omega_{p2}$=1 MHz. Solid line is the experimental result and the dotted line is the theoretical fitting curve corresponding to $P_{a_{-1}} = 32\%$, $P_{a_0} = 36\%$, and $P_{a_{+1}} = 32\%$. Other parameters are $\gamma_{ab}$=2 MHz, $\gamma_{ac}$=4 MHz, and $N_{F1}$=1.2×10¹¹ cm⁻³.

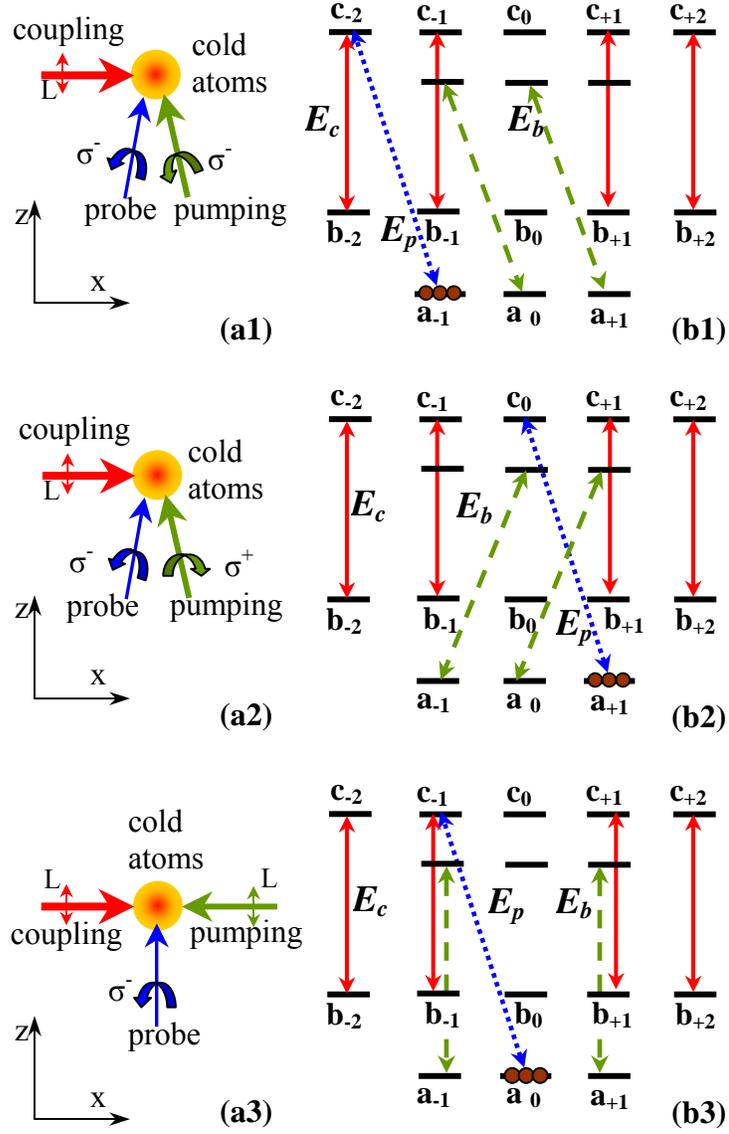

FIG. 2. (Color online) Different polarized pumping beam schemes. (a1)-(a3) Experimental setups for different laser beam propagation and polarization directions. (b1)-(b3) Relevant atomic energy levels with interacting laser beams. Solid (red) lines are coupling fields; dotted (blue) lines are probe fields, and dashed (green) lines are pumping fields.

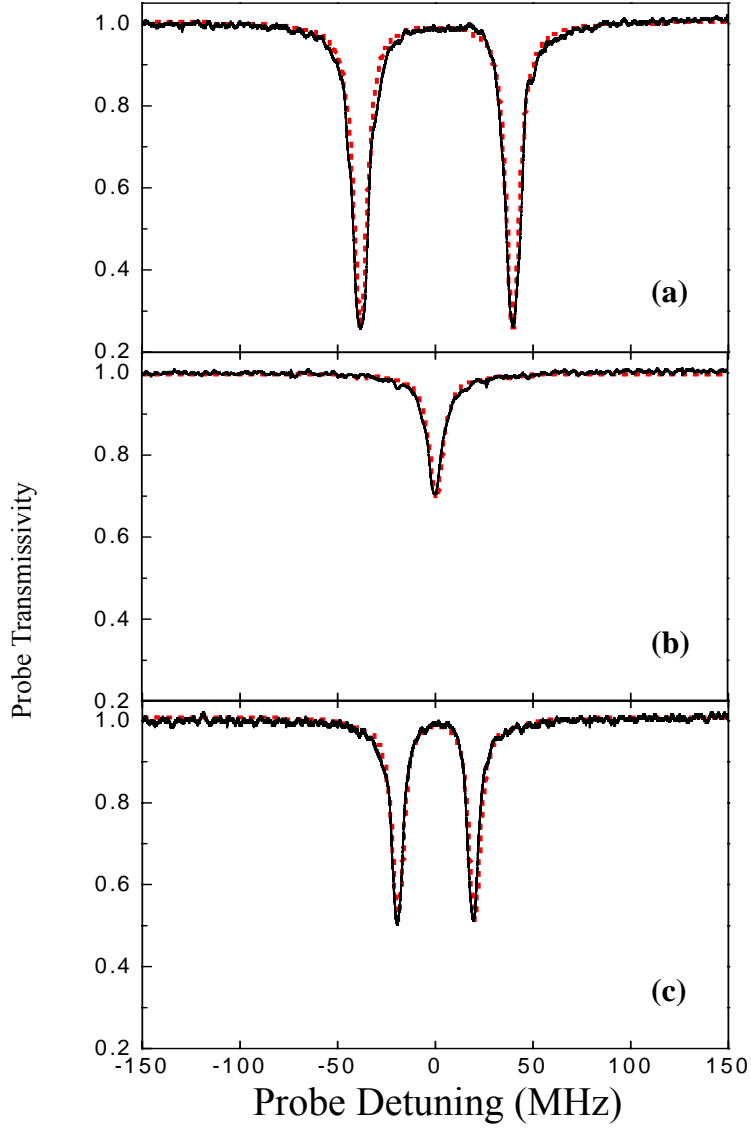

FIG. 3. (Color online) Transmission spectra of the probe beam for different pumping beam polarizations. Solid lines are the experimental results and dotted lines are the theoretical fitting curves corresponding to (a) left-circularly-polarized pumping beam with $P_{a_{-1}} = 96\%$, $P_{a_0} = 2\%$, $P_{a_{+1}} = 2\%$; (b) right-circularly-polarized pumping beam with $P_{a_{-1}} = 1\%$, $P_{a_0} = 1\%$, $P_{a_{+1}} = 98\%$; and (c) linearly-polarized pumping beam with $P_{a_{-1}} = 1\%$, $P_{a_0} = 98\%$, $P_{a_{+1}} = 1\%$. Other parameters are $\Omega_{c2}$=78 MHz, $\Omega_{p2}$=1MHz, $\gamma_{ab}$=2 MHz, $\gamma_{ac}$=4 MHz, and $N_{F1}$=0.6×10$^{11}$ cm$^{-3}$.

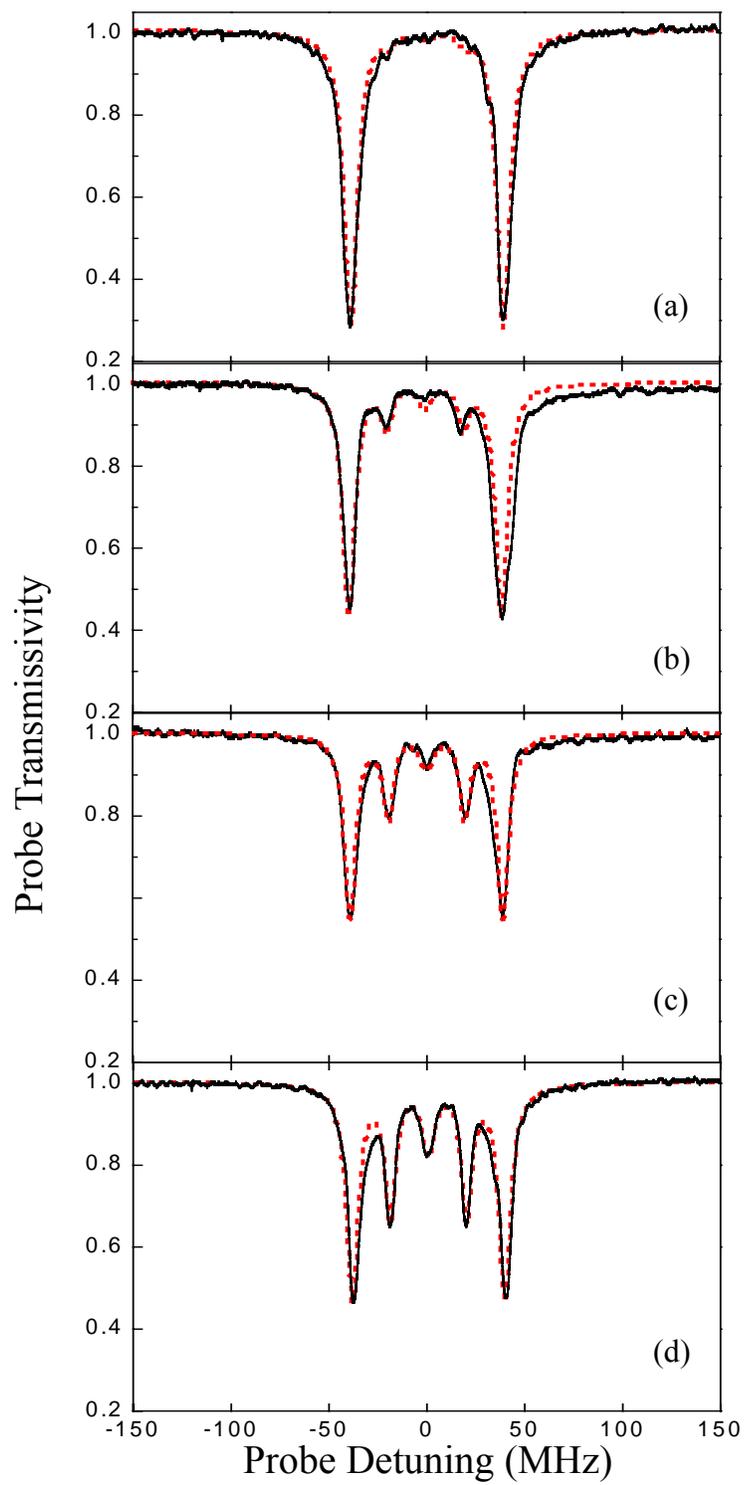

FIG. 4. (Color online) Transmission spectra of the probe beam for left-circularly-polarized pumping beam of different powers. Solid lines are the experimental results, and dotted lines are the theoretical fitting curves corresponding to (a) pumping beam power of 5 mW, $P_{a_{-1}} = 92\%$, $P_{a_0} = 4\%$, $P_{a_{+1}} = 4\%$, $N_{F1}=0.6\times10^{11}$ cm$^{-3}$; (b) pumping power of 1 mW, $P_{a_{-1}} = 64\%$, $P_{a_0} = 18\%$, $P_{a_{+1}} = 18\%$, $N_{F1}=0.6\times10^{11}$ cm$^{-3}$; (c) pumping power of 0.5 mW, $P_{a_{-1}} = 42\%$, $P_{a_0} = 33\%$, $P_{a_{+1}} = 25\%$, $N_{F1}=0.6\times10^{11}$ cm$^{-3}$; and (d) pumping power of 0.05 mW, $P_{a_{-1}} = 33\%$, $P_{a_0} = 35\%$, $P_{a_{+1}} = 32\%$, $N_{F1}=1\times10^{11}$ cm$^{-3}$.